\documentclass[aps,prl,twocolumn,floatfix,superscriptaddress]{revtex4-1}
\usepackage{amsmath}
\usepackage{graphicx}
\usepackage{color}
\usepackage{bm}
\usepackage{bigints}
\usepackage{comment}
\usepackage[colorlinks]{hyperref}
	\hypersetup{citecolor=red,urlcolor=blue}
\usepackage{blkarray}
\usepackage{tikz}
\usepackage{rotating}
\usepackage[export]{adjustbox}
\usepackage[caption=false,justification=justified]{subfig}
\usetikzlibrary{shapes,arrows,automata}
\usetikzlibrary{patterns}
\usetikzlibrary{intersections}

\newcommand{\beq}{\begin{equation}}
\newcommand{\eeq}{\end{equation}}

\newcommand{\degr}{\(^{\circ}\)}

\newcommand{\ctamop}{\affiliation{Centre for Theoretical Atomic, Molecular and Optical Physics, School of Mathematics and Physics, Queen's University Belfast,
\\
University Road, Belfast, BT7 1NN, Northern Ireland, UK}}

\newcommand{\tcd}{\affiliation{School of Physics and CRANN Institute, Trinity College Dublin, Dublin 2, Ireland}}

\newcommand{\charlesuni}{\affiliation{Institute of Theoretical Physics, Faculty of Mathematics and Physics, Charles University, V Hole\v{s}ovi\v{c}k\'{a}ch 2, 180 00 Prague 8, Czech Republic}}

\newcommand\hlight[1]{}

\newcommand\rhlight[1]{}

\newcommand\ghlight[1]{}

\begin{document}

\title{Enhancing Spin Polarization Using Ultrafast Angular Streaking}

\author{G. S. J. Armstrong}
\ctamop
\email[]{gregory.armstrong@qub.ac.uk}
\author{D. D. A. Clarke}
\tcd
\author{J. Benda} 
\charlesuni
\author{J. Wragg}
\ctamop
\author{A. C. Brown}
\ctamop
\author{H. W. van der Hart}
\ctamop

\date{\today}
\begin{abstract}
Through solution of the multielectron, semi-relativistic, time-dependent Schr\"{o}dinger equation, we show that angular streaking produces strongly spin-polarized electrons in a noble gas. The degree of spin polarization increases with the Keldysh parameter, so that angular streaking --- ordinarily applied to investigate tunneling --- may be repurposed to generate strongly spin-polarized electron bunches. Additionally, we explore modifications of the angular streaking scheme that also enhance spin polarization.
\end{abstract}


\maketitle

Photoelectron spin alignment represents an exploitable degree of freedom for probing both the structure and dynamics of matter.
Spin-polarized electron bunches facilitate the imaging of magnetic domains in surfaces \cite{altmanetal1991,rougemaille2010} and thin films \cite{Wuetal2013,Chenetal2015,Prietoetal2016}, as well as the study of magnetization profiles in magnetic nanostructures \cite{Dingetal2005,zdybetal2006,Niuetal2017}. Polarized electron beams are also instrumental in fundamental experiments addressing electron scattering and fluorescence emission from atoms and molecules \cite{Uhrigetal1994,Prinzetal1995,Greenetal2004,MasebergGay2009}, and even play a vital role in tests of the Standard Model with high-energy, electron-positron colliders \cite{vauth2016}.
Harnessing the spins of field-ionized electrons for real-time imaging and spectroscopy demands reliable and efficient experimental strategies for generating electron bunches with well-defined spin polarization. In this letter, we demonstrate 
numerically that the contemporary technique of attosecond angular streaking \cite{eckle2008,eckle2008sc} (or the ``attoclock"), traditionally used to investigate electron tunneling, can be repurposed to produce strongly spin-polarized electrons.

The production of spin-polarized electrons through the laser-driven ionization of atoms and molecules has been examined for over 50 years \cite{fano1969,lambrop1973,huang1979,dixit1981,nakajima2002}.
Their realization through the multiphoton ionization of noble gases by circularly polarized pulses has constituted an especially significant focus of experimental \cite{hartung2016,mmliu2018,trabert2018} and theoretical \cite{barth2013,barth2014} efforts alike.
In many of these studies, spin polarization was obtained in long circular pulses from the natural energy separation of corotating and counter-rotating ($p_{\pm1}$) electrons, which tend to have opposing spin orientations \cite{trabert2018}. In particular, low-energy photoelectron emission can be strongly spin-polarized, due to the suppression of emission of corotating electrons. In atoms with significant spin-orbit splittings, such as Xe, this effect is made more apparent: the energy separation between the ejected electrons enables resolution
of the final ionic ground-state level following ionization
\cite{mmliu2018,trabert2018}.
Diatomic molecules such as NO have also been identified as promising candidates for spin-polarized electron production \cite{liu2016,liu2017}, enabled by the opposing spin orientations of the valence $\pi_{\pm}$ orbitals, as well as their spectral separation. Strong-field schemes for producing spin-polarized electrons using orthogonal two-color pulses \cite{han2019} and bicircular pulses \cite{milosevic2016,ayuso2017,milosevic2018} have also been proposed.  

Short-pulse schemes such as angular streaking \cite{eckle2008,eckle2008sc} may provide an alternative means of tuning the spin polarization. The attoclock employs a few-cycle, near-circularly-polarized laser pulse of infrared wavelength to modify the binding potential of the target to form a rotating barrier, through which an electron may tunnel. The brevity of the pulse serves to localize the ejected-electron wavepacket within an angular interval, allowing a dominant emission direction to be defined. If emission is directed away from the major axis of the laser polarization ellipse, this could imply that electron ejection lags behind the laser field, by a time needed for tunneling.
Considerable controversy exists regarding this interpretation, with conflicting findings of tunneling being instantaneous \cite{eckle2008, eckle2008sc,pfeiffer2012,torlina2015,bray2018,sainadh2019} and non-instantaneous \cite{boge2013,landsman2014, camus2017}. A detailed review on this subject is given in Ref.\;\cite{kheifetsatto}.

Regardless of the attoclock's ability to provide a tunneling time, 
calculations have shown that in an attoclock scheme, $p_{\pm 1}$ electrons ejected from noble gases emerge at different angles to the major axis of the laser polarization, and are thereby spatially separated to some extent \cite{kaushal2015,liu2018}. These electrons are also separated in energy \cite{barth2011} and favor opposite spin orientations
\cite{trabert2018}, such that the differential in both energy and emission angle may render the attoclock scheme promising for the realization of 
spin-polarized electron bunches.


\begin{figure}[t]
    \centering
    \includegraphics[width=\columnwidth]{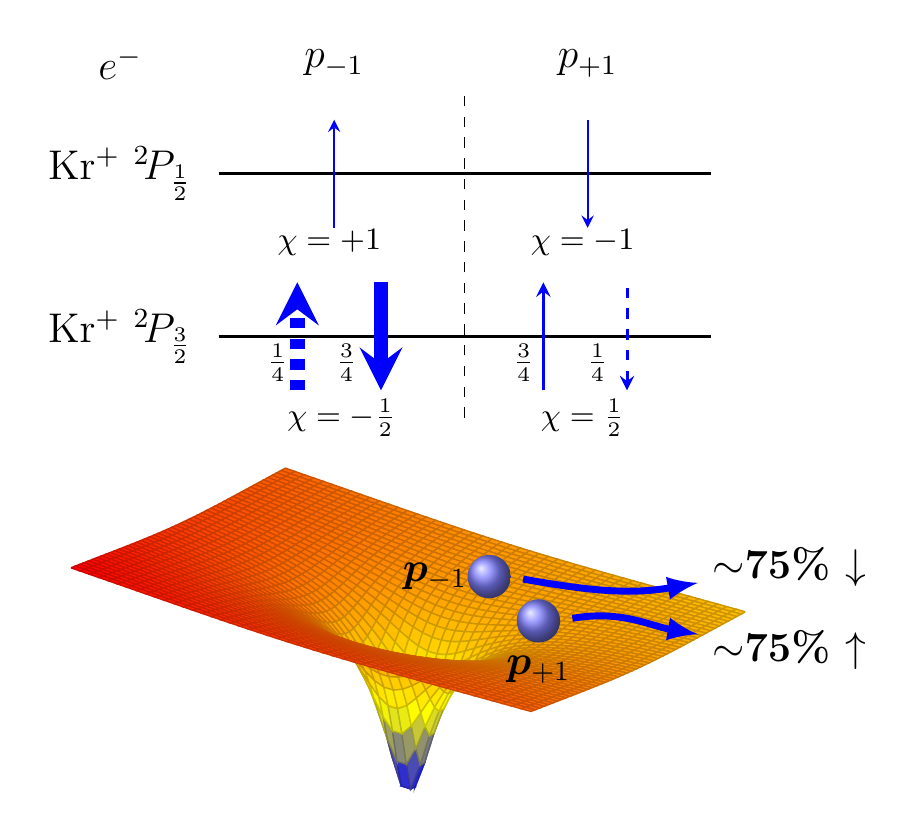}
    \caption{Schematic diagram of ionization pathways for Kr. 
    Upper panel: arrows indicate spin orientation, and their thickness indicates the relative yield. 
    Lower panel: The $p_{\pm 1}$ electrons ejected in the attoclock scheme emerge with different momentum vectors, and their differing spin preferences enables spin polarization. Values of the spin polarization $\chi$ are indicated. See text for further details.}
    \label{schematic}
\end{figure}

Figure \ref{schematic} illustrates this separation of co- and counter-rotating electrons
in the angular streaking of the Kr atom. Note that in this work, we define the electric field such that the $p_{-1}$ electron is counter-rotating, and the $p_{+1}$ electron corotating.
An elliptically-polarized, few-cycle pulse ejects valence $4p_{\pm1}$ electrons, leaving the residual Kr$^+$ ion in either the $^2P_{3/2}$ ground state, or the $^2P_{1/2}$ excited state,
with a fine-structure splitting of about 0.67 eV. 
As shown in Fig.\;\ref{schematic}, the $^2P_{3/2}$ ionic state is fourfold degenerate, coupling to a $p_{-1}$ electron that prefers to be spin down, and a $p_{+1}$ electron that prefers to be spin up \cite{trabert2018}. The $^2P_{1/2}$ excited state is twofold degenerate: the $p_{+1}(p_{-1})$ electron must be spin down(up). This degeneracy imposes limits on the highest possible degree of spin polarization: if the ion is left in the $^2P_{1/2}$ excited state, spin polarization (given by the $\chi$ values indicated in Fig.\;\ref{schematic}) can reach 100\%, whereas if it is left in the $^2P_{3/2}$ ground state, the spin polarization can attain a maximum magnitude of 50\%. Counter-rotating electrons dominate the yield in strong-field ionization, and hence the strongest spin polarization to be expected is --50\%. As shown in the lower panel of Fig.\ref{schematic}, the spatial separation of $p_{\pm1}$ electrons enabled by the angular streaking scheme \cite{kaushal2015} means that strong spin polarization may be obtained in certain spatial directions.



We explore spin polarization driven by angular streaking of Kr using the {\em ab initio}, $R$-matrix with time-dependence (RMT) method \cite{nikolopoulos2008,moore2011,clarke2018,rmtcpc}. RMT provides a solution of the time-dependent Schr\"{o}dinger equation in full dimensionality for single ionization of multielectron atomic and molecular systems.
The atomic structure input for the RMT calculations is obtained from both non-relativistic and semi-relativistic calculations using the \textsc{rmatrx i} codes \cite{berrington1995}. Our atomic structure model of the Kr atom is described in Ref.\;\cite{wragg2019}. We use bound orbitals obtained from Hartree-Fock calculations \cite{clemroe}, and allow the ejection of both $4s$ and $4p$ electrons. 



The time-varying laser field is treated classically and within the electric dipole approximation. The $\hat{z}$-axis is along the laser propagation direction, so that the field is constrained to the $xy$-plane. Thus, we have
\begin{equation}
    \bm{\mathcal E}(t) = \frac{{\cal E}_0}{\sqrt{1+\epsilon^2}} \sin^4 \left(\frac{\omega t}{2N_c}\right)
    \left[
        \cos\omega t \;\hat{\bf x} + \epsilon\sin\omega t \;\hat{\bf y}
    \right],
\end{equation}
where \({\cal E}_0\) is the peak electric field strength,
\(\omega\) is the laser frequency, \(N_c\) is the number of laser cycles, and \(\epsilon\) is the ellipticity. The field retains this profile for time $t \in [0,2\pi N_c/\omega]$, and is zero otherwise. The pulses considered in this work ramp on over four cycles, followed by an equal number of cycles of ramp-off, so that \(N_{c} = 8\) in all cases.
We propagate the wave function using an Arnoldi propagator of order 8, 
for a total of 60 fs.
The ejected electron is described up to a distance of $3730\ a_{0}$ from the nucleus, where $a_{0}$ is the Bohr radius.

Following time propagation, we obtain the photoelectron momentum distribution in
the laser polarization plane. The contributions of co- and counter-rotating electrons, as well as both spin-up and spin-down electrons, are resolved by decoupling the spin and orbital angular momenta of the ejected electron from those of the residual Kr\(^+\) ion. The ejected electron wavefunction is transformed to momentum space using a Fourier transform.

\begin{figure}[t]

        \subfloat[\label{corot} Corotating ($p_{+1}$)]
        {\includegraphics[width=0.47\columnwidth,clip=true,trim={1cm 2cm 2cm 1cm}]{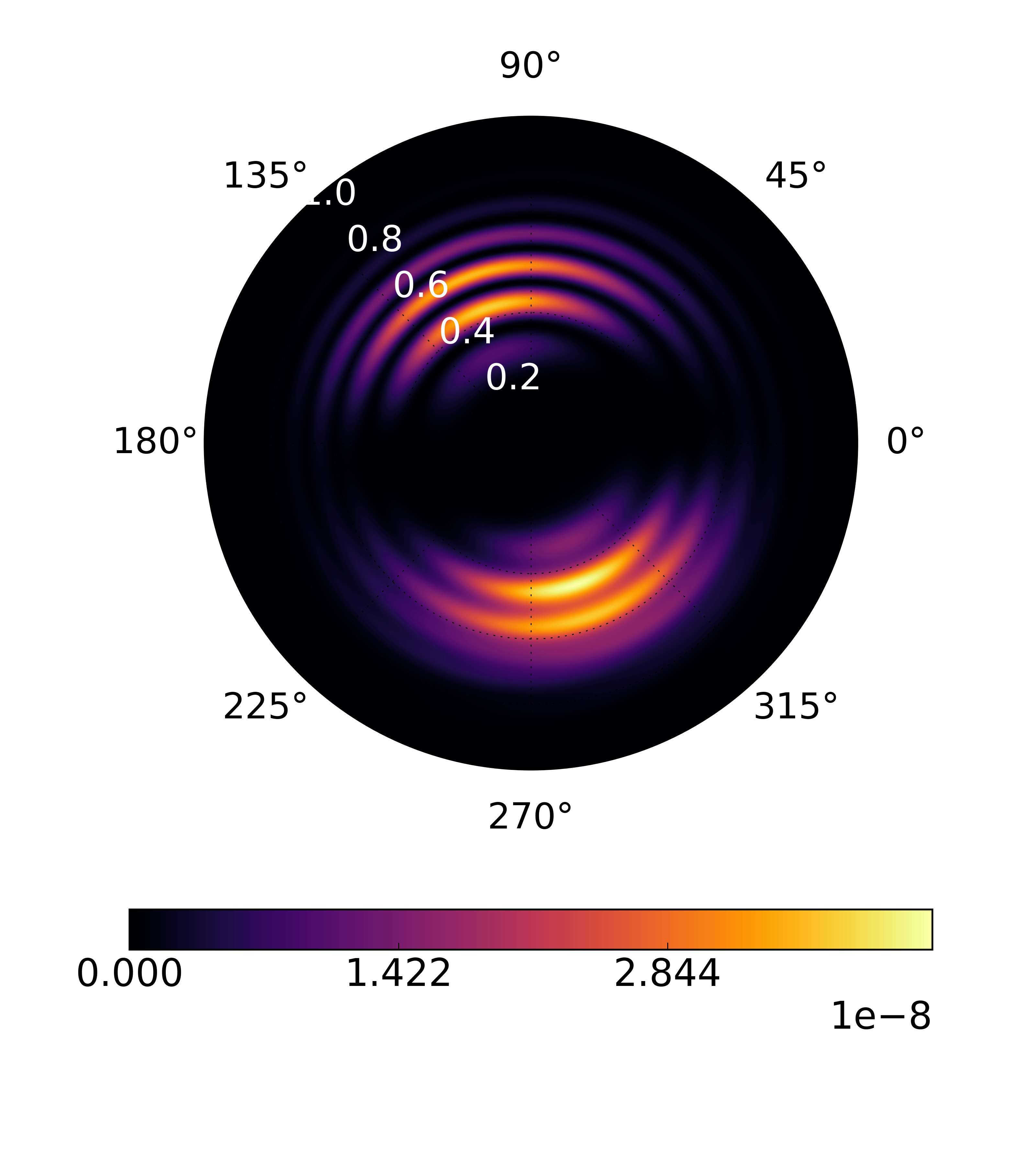}}
    \quad
        \subfloat[\label{counterrot} Counter-rotating ($p_{-1}$)]
        {\includegraphics[width=0.47\columnwidth,clip=true,trim={1cm 2cm 2cm 1cm}]{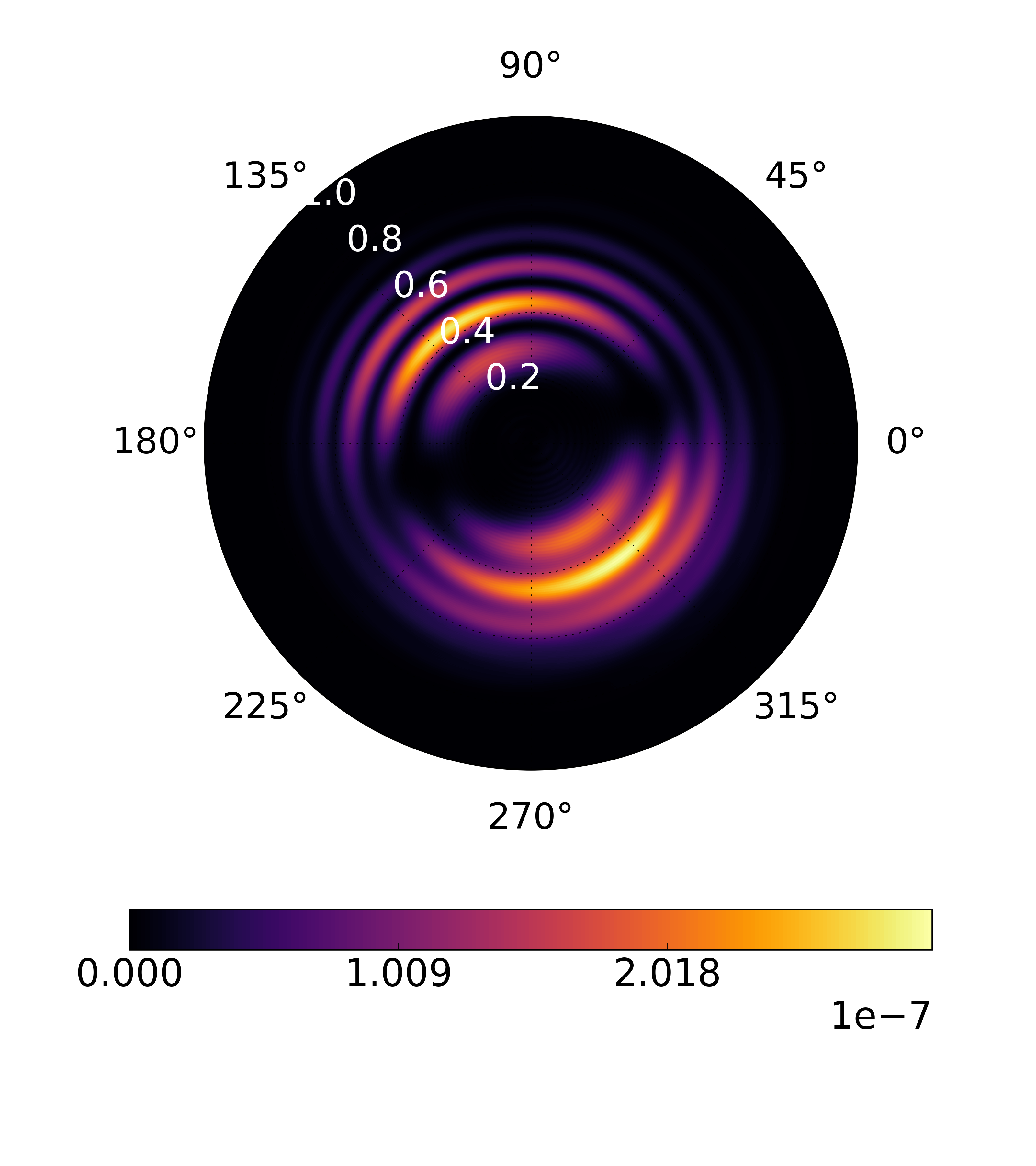}}

    \vskip\baselineskip

        \subfloat[\label{lsangdis}]
        {\includegraphics[width=0.94\columnwidth]{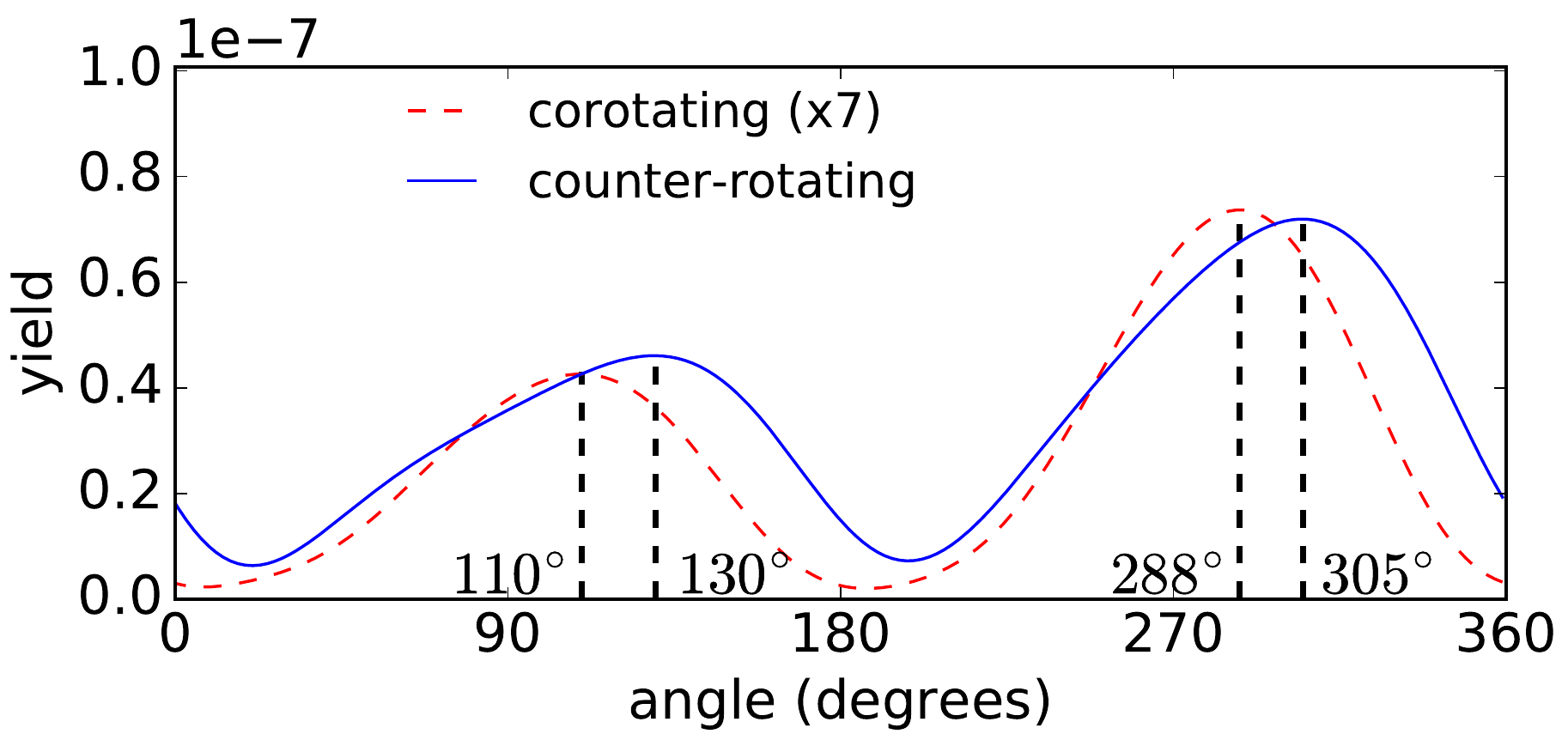}}
    \hfill
    
    \caption{Momentum distribution for (a) corotating ($p_{+1}$) and (b) counter-rotating ($p_{-1}$) electrons ionized from Kr by an 8-cycle, 780-nm, 2\(\times10^{13}\) W/cm\(^2\) pulse of ellipticity \(\epsilon=0.87\). The radial labels indicate momentum values (in a.u.). Note the different orders of magnitude in the scales of (a) and (b). (c) The respective momentum-integrated distributions.}
    \label{totals}
\end{figure}

We first investigate the laser-driven response of Kr in a conventional angular streaking scheme. Fig.\;\ref{totals} shows the polarization-plane, photoelectron momentum distributions for Kr, ionized by an 8-cycle, $2\times10^{13}$ W/cm$^2$, elliptically polarized (\(\epsilon=0.87\)), 780-nm pulse, separated into contributions from co- ($p_{+1}$) and counter-rotating ($p_{-1}$) electrons (Figs.\;\ref{corot} and \ref{counterrot} respectively). The elliptically polarized laser pulse confines the dominant ionization yield to an angular range, in contrast to the approximately angle-independent yield induced by circular pulses. The momentum-integrated distribution in Fig.\;\ref{lsangdis} demonstrates a clear angular separation of around 10\degr\ between co- and counter-rotating electrons. Such separation is typically attributed to differing deflections caused by the Coulomb potential \cite{kaushal2015,liu2018}.
Counter-rotating electrons provide close to 90\% of the total yield.

Intuitively, the strongest spin polarization should be attained if the yield of corotating electrons can be minimized. This is naturally achieved at low photoelectron energies \cite{mmliu2018,trabert2018,barth2013}. However, angular streaking provides an additional source for this minimization, by confining the dominant photoelectron yield to a limited angular range, thereby creating the minima seen in Fig.\;\ref{lsangdis}. Spin polarization should then reach an optimal value close to \(\phi = 0, 180\)\degr, where the corotating yield is minimal.

\begin{figure}[t!]
        \subfloat[\label{momspinup735} Spin up]
        {\includegraphics[width=0.47\columnwidth,clip=true,trim={1cm 2cm 2cm 1cm}]{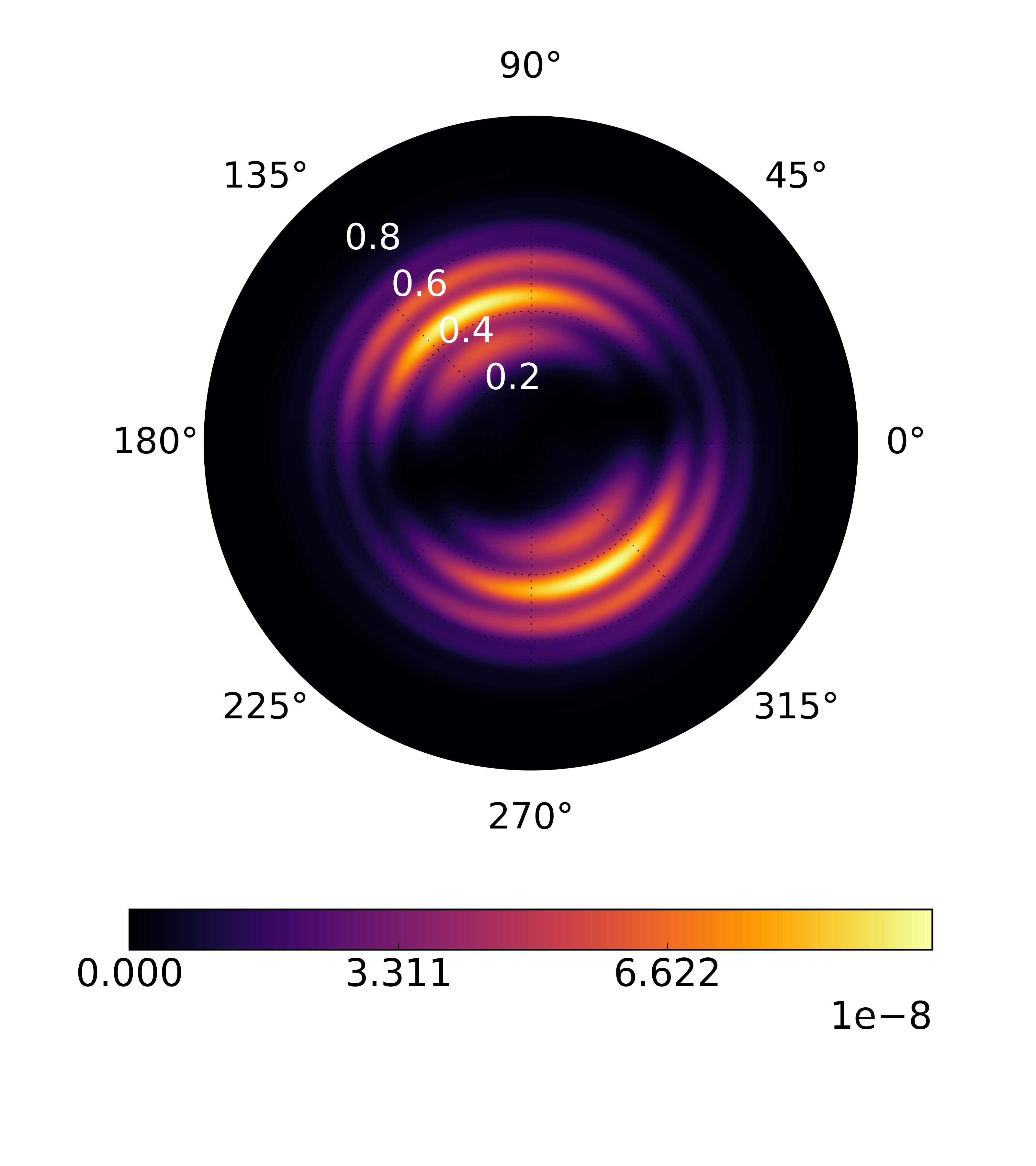}}
    \quad
        \subfloat[\label{momspindown735} Spin down]
        {\includegraphics[width=0.47\columnwidth,clip=true,trim={1cm 2cm 2cm 1cm}]{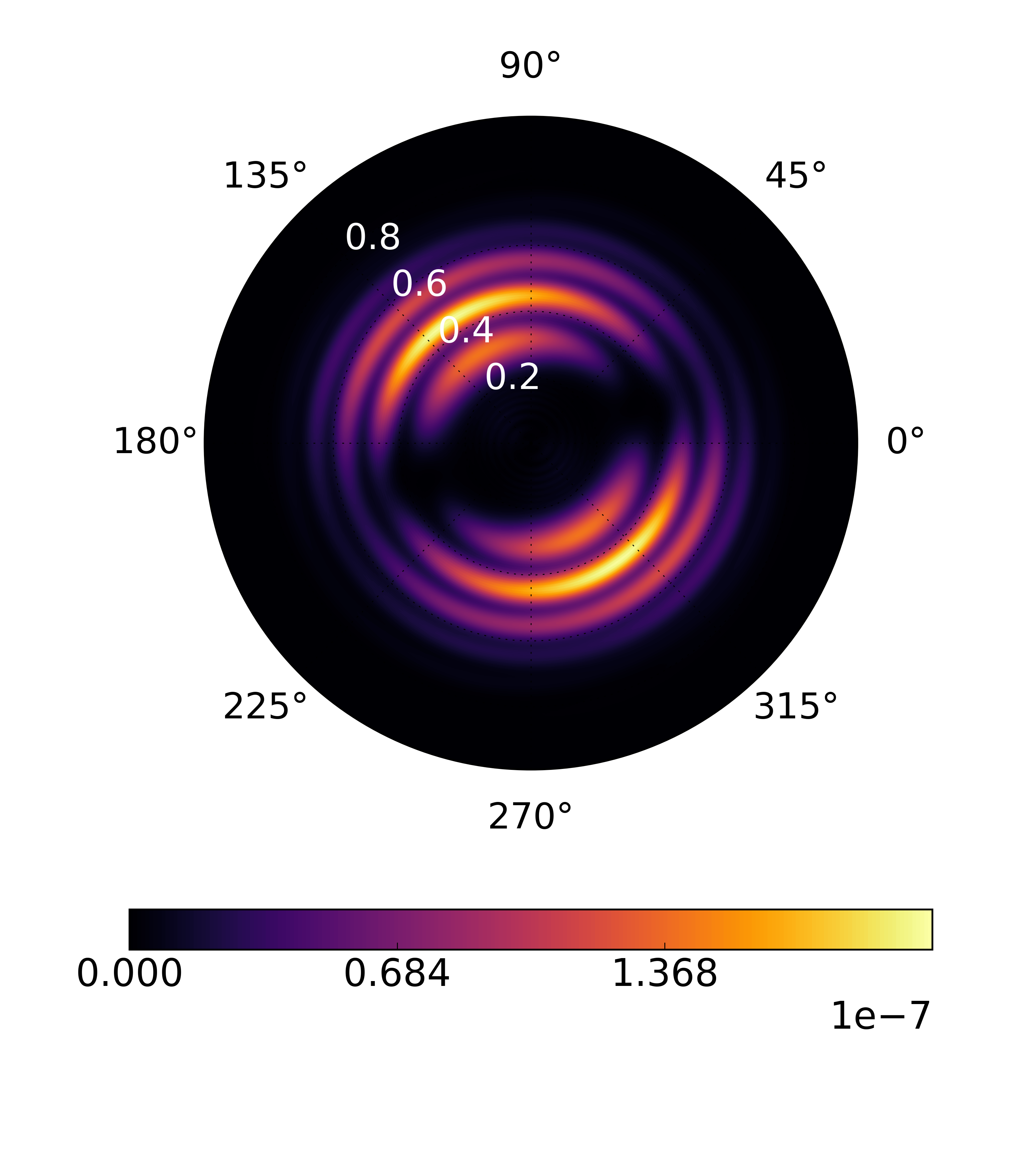}}
        \vspace{0.1in}
        \subfloat[\label{kint735r}]
        {\includegraphics[width=0.94\columnwidth]{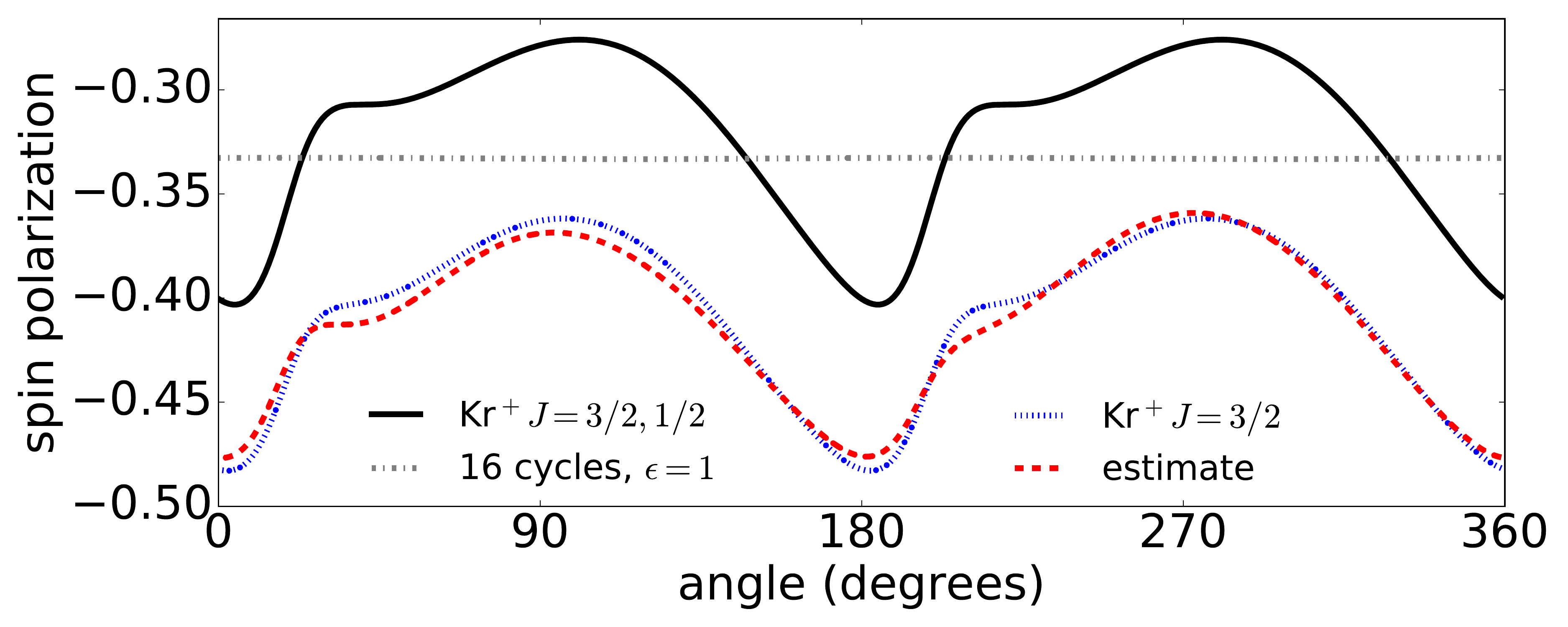}}
        
    \caption{Momentum distribution for (a) spin up and (b) spin down electrons ionized from Kr by an 8-cycle, 780-nm, 2\(\times10^{13}\) W/cm\(^2\) pulse of ellipticity \(\epsilon=0.87\). Note the different orders of magnitude in the color scales.
    (c) Momentum-integrated spin polarization, its dependence on the Kr$^+$ $J$ value, and its single-threshold ($J=3/2$) estimate.  
    }
    \label{krspins}
\end{figure}

To investigate this, we calculate the spin-resolved momentum distributions for photoelectrons ejected from Kr in the same 
scheme as in Fig.\;\ref{totals}. Figs.\;\ref{momspinup735} and \ref{momspindown735} show the polarization-plane momentum distributions for spin-up and spin-down electrons. The energy- and angle-dependence of both distributions show a clear resemblance to the distribution in Fig.\;\ref{counterrot}, which is not surprising, given that counter-rotating electrons dominate the overall yield. 
Their dominance also explains the observation that spin down electrons are strongly favored.

To assess the angular variation of spin polarization in greater detail, we integrate the spin-resolved momentum distributions over momentum, and define the angular distributions of spin-up  and spin-down  electrons as ${\cal P}_{\uparrow}(\phi)$ and ${\cal P}_{\downarrow}(\phi)$ respectively. From these quantities, we obtain an angle($\phi$)-dependent spin polarization given by
$    \chi(\phi) = 
[{\cal P}_{\uparrow}(\phi) - {\cal P}_{\downarrow}(\phi)]
/
[{\cal P}_{\uparrow}(\phi) + {\cal P}_{\downarrow}(\phi)]
.
$
Fig.\;\ref{kint735r} shows this quantity as a function of azimuthal angle (solid black curve). A significant spin polarization of around $-0.4$ is attained close to $\phi=0,180$\degr, as expected (see Fig.~\ref{lsangdis}). Away from these angles, corotating electrons make stronger, though still small, contributions.
Since corotating electrons prefer to be spin up in this case, their contribution weakens the total spin polarization. Indeed, the weakest spin polarization observed here, of around $-0.28$, is attained close to \(\phi = 100,280\)\degr, where the corotating electron yield is maximal. At other angles, the spin polarization varies, since it comprises angle-dependent combinations of co- and counter-rotating electron yields. 
The angle dependence of the spin polarization is related to the angle-dependent ratio of counter-rotating to corotating electron yields, $R(\phi)$. Considering only the ionic ground state, and assuming the spin weights indicated in Fig.~\ref{schematic}, 
a simple, single-threshold estimate for the spin polarization
can be obtained via $\chi \approx -\frac{1}{2}[R(\phi) -1]/[R(\phi)+1]$. We plot this estimate in Fig.\;\ref{kint735r}, alongside the spin polarization calculated using the RMT method. Thus, an angular separation of co- and counter-rotating electrons which produces a large value of $R(\phi)$ ought to yield strong spin polarization.

The calculated spin polarization is significantly weaker than this estimate, 
due to contributions from the $J = 1/2$ ionic state which are strongly spin up. As shown in Fig.\;\ref{kint735r}, when we include only the ground ionic state in our analysis, the calculated spin polarization agrees well with the
simple estimate. Under these conditions, the ionic excited state contributes around 8\% of the total yield.

Of course, in a strong-field scheme some degree of spin polarization is likely to be observed, due to the natural imbalance between co- and counter-rotating electron yields. This is demonstrated in recent experiments \cite{trabert2018,mmliu2018}, which use long circular pulses to obtain an energy-dependent spin polarization. To investigate the particular effectiveness of the angular streaking scheme, we show in Fig.\;\ref{kint735r} the spin polarization calculated using a 16-cycle, circularly polarized pulse ($\epsilon=1$). Relative to this scheme, angular streaking clearly enhances the spin polarization, yielding a minimum spin polarization which is around 20\% lower than the angle-independent value of around --0.33 obtained using a long circular pulse, equivalent to around 40\% progress towards the limit of --0.5.
{\color{black}The pronounced angular minimum in the corotating electron yield created by the attoclock facilitates its advantage over the long-pulse scheme, conferring a higher degree of spin polarization and dominant emission in a well defined angular interval.}

\begin{figure}[t]
    \centering
    \includegraphics[width=0.7\columnwidth]{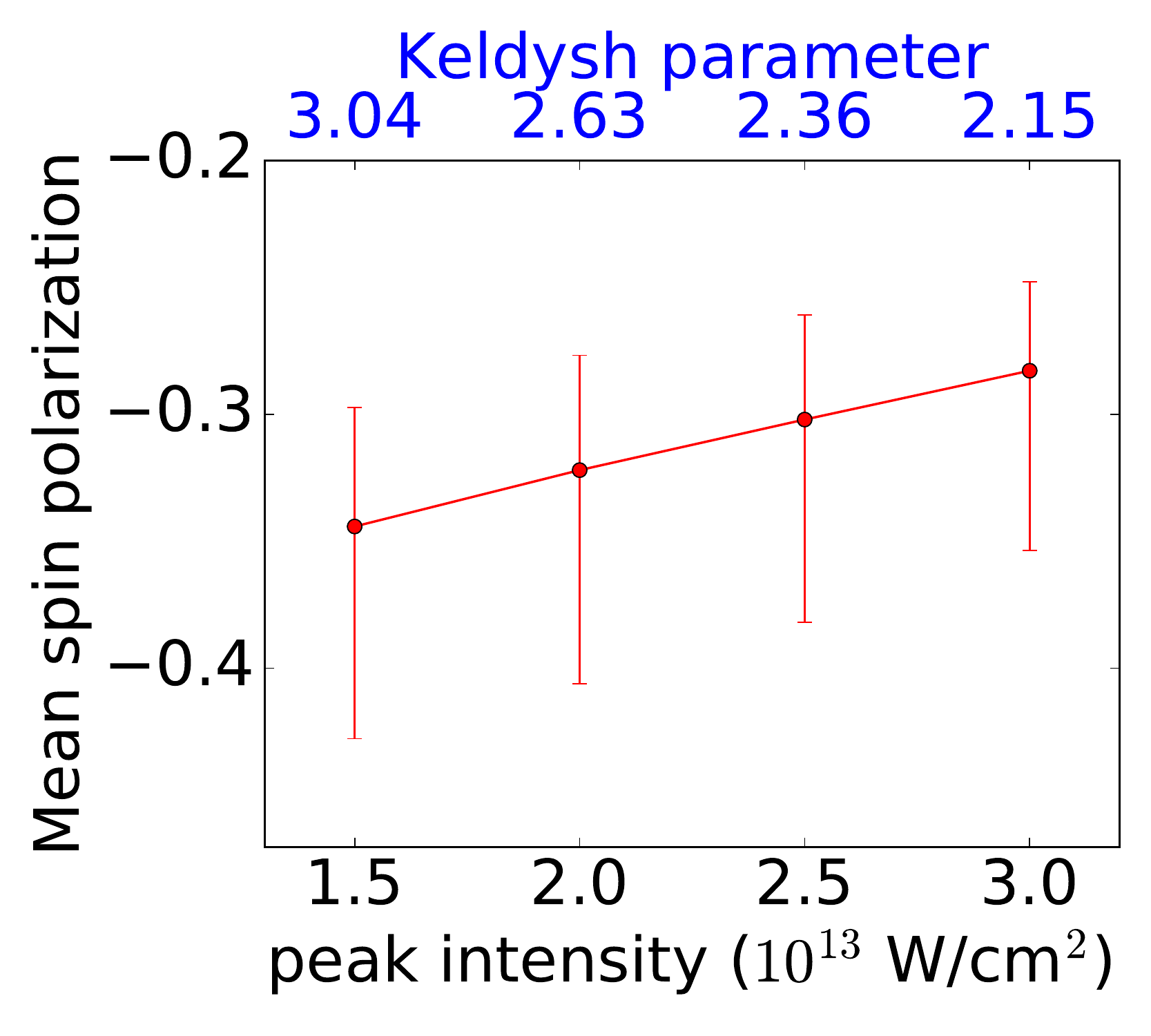}
    \caption{Average energy-integrated spin polarization as a function of laser intensity. Vertical bars indicate the range of spin polarization values obtained over the full range in azimuthal angle $\phi$.}
    \label{idep735nm}
\end{figure}

Although a strong spin polarization is obtained under these conditions, the total photoelectron yield is rather low. Although the yield may be enhanced by increasing the peak laser intensity, this would be detrimental to the spin polarization, since corotating electron contributions would increase. 
This effect is made clear in Fig.\;\ref{idep735nm}, where we show the calculated mean value of the energy-integrated spin polarization as a function of peak laser intensity. The vertical bars indicate the range of values taken by the spin polarization over all azimuthal angles. Over this range of intensities, the average spin polarization reduces in magnitude by around 15\%, as do the extreme (most and least negative) values. 


The need for a relatively low peak intensity for the purpose of spin polarization contrasts with the typical attoclock requirement of relatively high peak intensity to access the tunneling regime (note the Keldysh parameter values in Fig.~\ref{idep735nm}). This indicates that the attoclock scheme may be applied to good effect in the multiphoton regime to generate strong spin polarization. We note also that since a low intensity is advantageous for spin polarization, focal-volume averaging is likely to enhance its degree. Therefore, we expect that strong spin polarization should be detectable in low-intensity experiments at a wavelength of 780 nm.
 

\begin{figure}[t]

        {\includegraphics[width=0.98\columnwidth,clip=true,trim={0 0 0 0}]{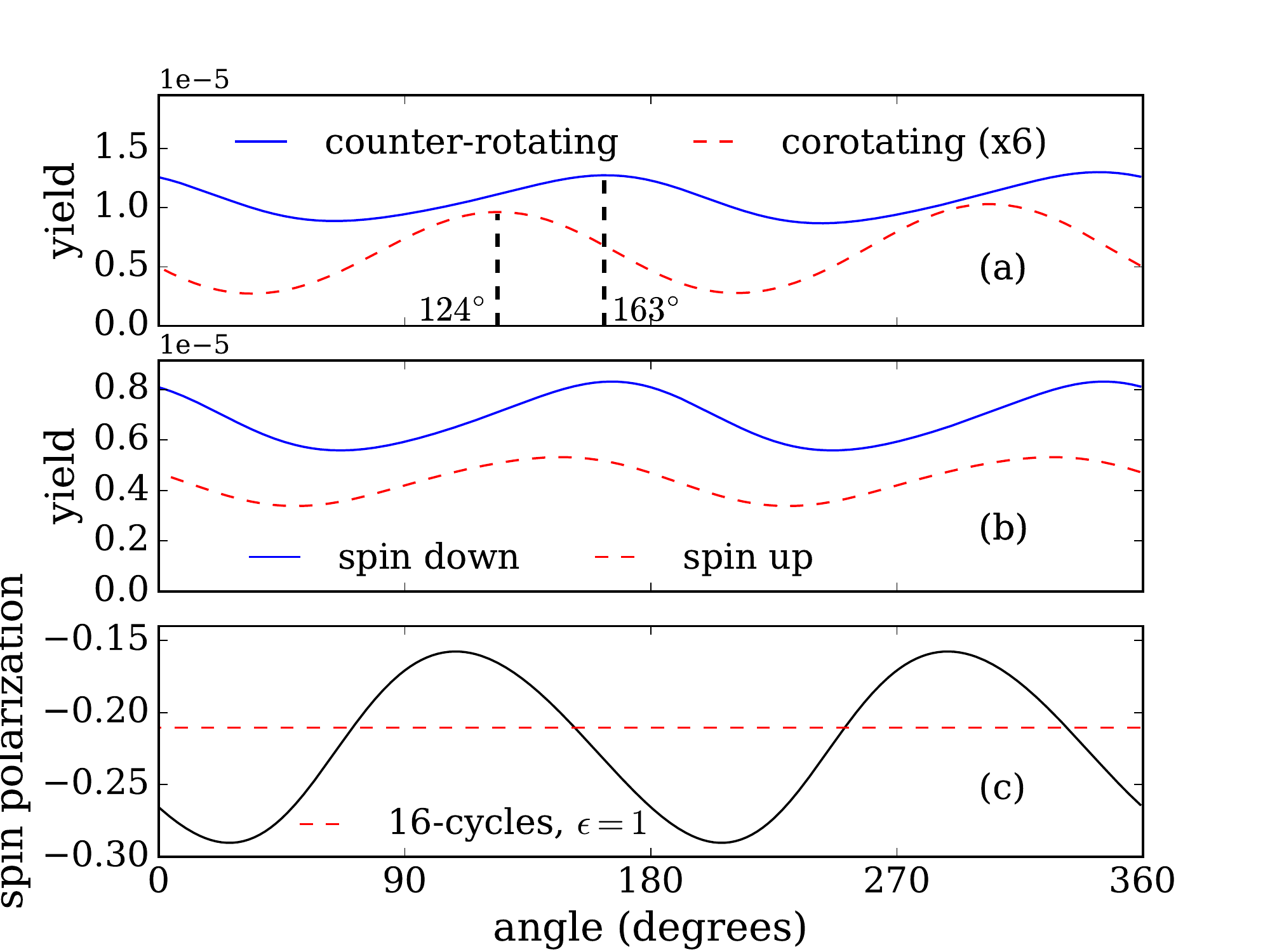}}
    \caption{Angular distributions for (a) corotating and  counter-rotating electrons, (b) spin-up and spin-down electrons, and (c) spin polarization of electrons ionized from Kr by an 8-cycle, 390-nm, 2\(\times10^{13}\) W/cm\(^2\) pulse of ellipticity \(\epsilon=0.87\).}
    \label{krspins390}
\end{figure}



Of course, it would be useful if strong spin polarization could be obtained by aligning the minimum in the corotating electron signal with a strong counter-rotating electron signal. Such a separation may be possible at shorter wavelengths, where the reduction in ejected-electron angular momentum ought to increase the angle between corotating and counter-rotating electrons. 

To do this, we choose a wavelength of 390 nm, and retain an ellipticity of 0.87, and a peak laser intensity of $2\times10^{13}$ W/cm$^2$ (as in Fig.\;\ref{krspins}). 
This wavelength may be viable in experiment, merely requiring frequency-doubling within the typical attoclock setup. Fig.\;\ref{krspins390}(a) shows the calculated angular distributions for corotating and counter-rotating electrons. The angular separation of 39$^{\circ}$ is significantly larger than that observed in Fig.\;\ref{totals} at 780 nm, and significant counter-rotating electron yield is observed at angles where the yield of corotating electrons is minimal. The photoelectron yield is also considerably larger than that obtained at 780 nm. Fig.\;\ref{krspins390}(b) shows the calculated angular distributions for spin-up and spin-down electrons. Spin-down electrons contribute almost twice the yield of spin-up electrons. The angle-dependent spin polarization, shown in Fig.\;\ref{krspins390}(c), reaches a value of $-0.3$ at around 200\degr, close to the minimum in the corotating-electron signal. 
Although this is a smaller degree of spin polarization than that obtained in the 780-nm case at the same peak intensity, its angular location lies in a region of relatively high photoelectron yield, and  strong values span a broader angular range. {\color{black}The main reason for the lower degree of spin polarization is that the minimum in the corotating electron yield is not as pronounced as that obtained at 780 nm (see Fig.\;\ref{totals}).}

 In Fig.\;\ref{krspins390}(c), we compare the spin polarization obtained using the angular streaking scheme with the result for a 16-cycle, circular pulse ($\epsilon=1$). As at 780 nm (see Fig.\;\ref{kint735r}), we find a significant enhancement of the spin polarization in the angular streaking scheme. In this case the enhancement is around 30\%, with the 16-cycle pulse yielding a spin polarization of around --0.21 at all angles. 

\begin{figure}[t]
    \centering
    \includegraphics[width=\columnwidth]{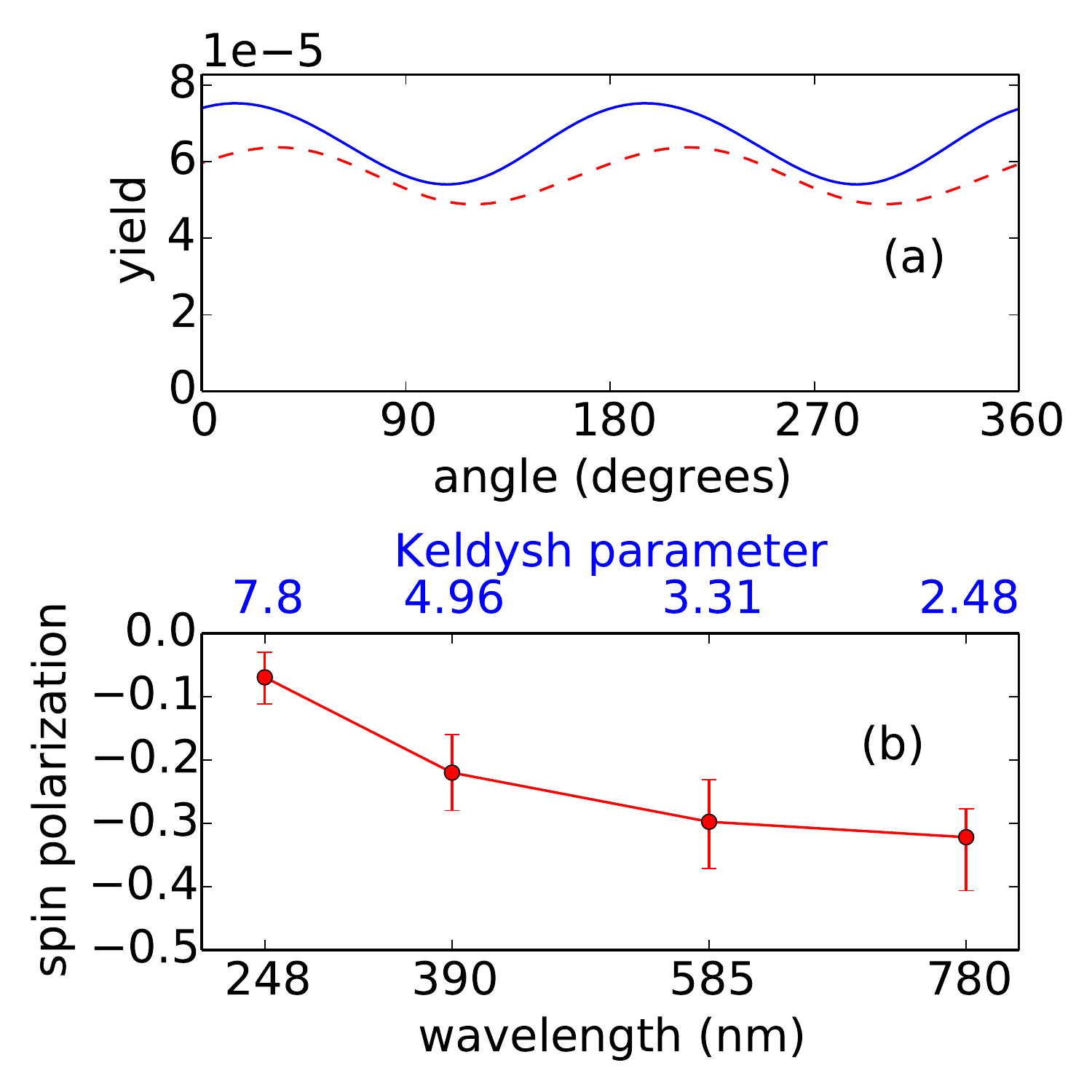}
    \caption{
    (a) Angular distribution of spin-up and spin-down electrons ionized from Kr by an 8-cycle, 248-nm, $2\times10^{13}$ W/cm$^2$ pulse of ellipticity $\epsilon=0.87$. 
    (b) Average energy-integrated spin polarization as a function of laser wavelength at an intensity of $2\times10^{13}$ W/cm$^2$. Vertical bars indicate the range of spin polarization values obtained over the full range in azimuthal angle $\phi$.}
    \label{wdep}
\end{figure}

As was the case at 780 nm, we find that spin polarization is reduced in magnitude as peak laser intensity is increased. We also find that further shortening of the wavelength is detrimental to spin polarization, since {\color{black}the minimum in the corotating electron yield is not as deep as that observed at 780 nm.}
Fig.\;\ref{wdep}(a) shows that if the laser wavelength is reduced to 248~nm (with the laser intensity fixed at $2\times10^{13}$ W/cm$^2$), spin-up  and spin-down electrons contribute on a similar level, in stark contrast to the dominance of spin-down electrons seen in Fig.\;\ref{krspins} and Fig.\;\ref{krspins390}(b). Consequently, the energy-integrated spin polarization approaches zero as laser wavelength is decreased to 248 nm, as shown in Fig.\;\ref{wdep}(b). As the wavelength is increased to 585 nm and beyond, the spin polarization appears to reach a plateau. This indicates that even with longer wavelengths, it may be difficult to achieve a spin polarization significantly larger than that obtained at 780 nm. Therefore, the parameters chosen in this work appear to be close to optimal for the generation of spin-polarized electrons within a single-pulse angular streaking scheme.


Our findings are not limited to atomic krypton. The angular streaking scheme should enhance spin polarization for any atomic or molecular system in which spin polarization can be generated by long, circularly polarized pulses. The streaking scheme introduces angular separation between co-rotating and counter-rotating electrons, and thus enhances the spatial separation of interfering pathways. In addition, the scheme has the advantage that it is robust over focal-averaging. As the laser intensity decreases, the degree of spin polarization increases. Hence, the angular streaking scheme should be applicable to a wide range of systems.

In conclusion, we have investigated the use of angular streaking as a generator of spin-polarized electrons. We have shown that the angular streaking scheme can induce spin polarization by minimizing the corotating electron yield over an angular range, thereby creating a region of strongly spin-polarized electrons. Additionally, we have shown that the angular streaking scheme may be used for this purpose when implemented in the low-intensity, multiphoton regime, far from the tunneling conditions under which the scheme is usually employed. We suggest this as a new application of angular streaking, from which the useful product of spin-polarized electrons may be obtained.

 The data presented in this article may be accessed at Ref.\;\cite{pure}. The RMT code is part of the UK-AMOR suite, and can be obtained for free at Ref.\;\cite{repo}. This work benefited from computational support by CoSeC, the Computational Science Centre for Research Communities, through CCPQ. The authors acknowledge funding from the UK Engineering and Physical Sciences Research Council (EPSRC) under grants EP/P022146/1, EP/P013953/1, EP/R029342/1, and EP/T019530/1. This work relied on the ARCHER UK National Supercomputing Service (\url{www.archer.ac.uk}), for which access was obtained via the UK-AMOR consortium funded by EPSRC.

\bibliography{mainbib}

\end{document}